\documentclass[aps,twocolumn]{revtex4}

\usepackage{amsmath,amssymb}
\usepackage{graphicx}
\newcommand{\amp}{&\!\!}

\begin{document}

\title{Some Calculable Contributions to Entanglement Entropy}

\author{Mark~P.~Hertzberg$^{1,2*}$ and Frank Wilczek$^{1}$}
\affiliation{$^1$Center for Theoretical Physics, Massachusetts Institute of Technology, Cambridge, MA 02139, USA\\
$^2$KIPAC and SITP, Stanford University, Stanford, CA 94305, USA}

\begin{abstract}
Entanglement entropy appears as a central property of quantum systems in broad areas of physics. However, its precise value is often sensitive to unknown microphysics, rendering it incalculable.
By considering parametric dependence on correlation length, we extract finite, calculable contributions to the entanglement entropy for a scalar field between the interior and exterior of a spatial domain of arbitrary shape.  The leading term is proportional to the area of the dividing boundary; we also extract finite subleading contributions for a field defined in the bulk interior of a waveguide in 3+1 dimensions, including terms proportional to the waveguide's cross-sectional geometry; its area, perimeter length, and integrated curvature. We also consider related quantities at criticality and suggest a class 
of systems for which these contributions might be measurable.
\end{abstract}

%\vspace*{-\bigskipamount} \preprint{MIT-CTP-4110} 

%\date{\today}

\maketitle

\begin{figure}[t]
\includegraphics[width=4.1cm]{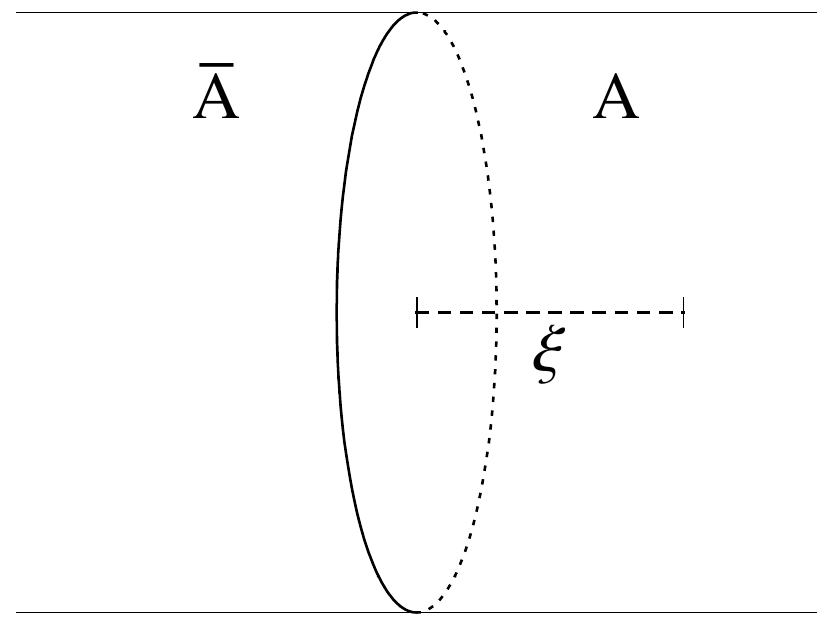}\,\,\,\,\,
\includegraphics[width=4.1cm]{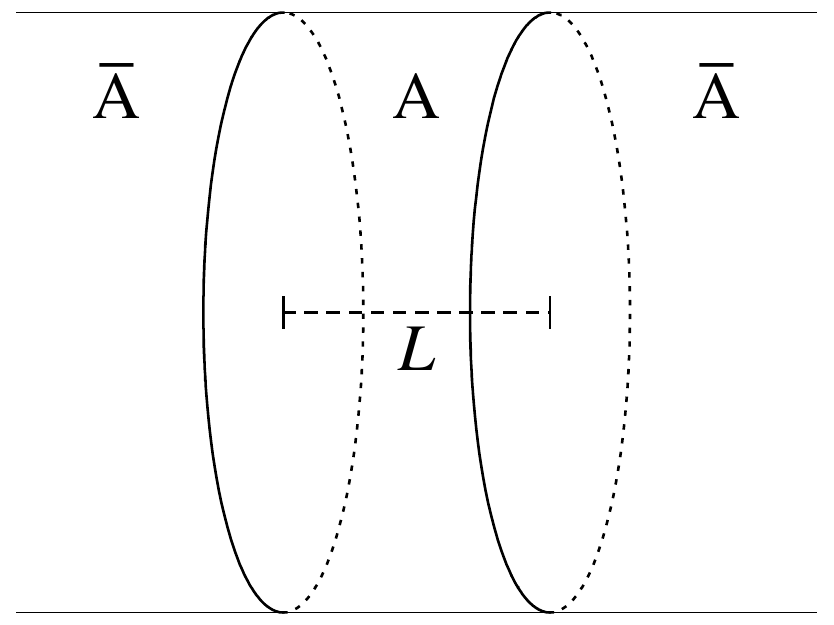}
\caption{Waveguide geometry in $d=3$. Left: Region A is a half-space at finite correlation length $\xi$. 
Right: Region A is an interval of length $L$ at criticality.}
\label{Waveguide}
\end{figure}

{\em Introduction}:
The quantum nature of matter is rarely evident on macroscopic scales, often due to the decoherence of excited states toward classical states. However, for certain states, such as ground/vacuum states, their quantum nature can appear, in principle, on macroscopic scales. One of the most dramatic properties of quantum matter is {\em entanglement} and its associated {\em entropy}, which, if observed on mesoscopic or macroscopic scales, would be of broad interest.
To define this entropy, consider a quantum system whose degrees of freedom can be divided into two parts in space A, $\overline {\rm A}$ (see Fig.~\ref{Waveguide}).   The geometric or entanglement entropy is defined by the von Neumann formula  $S=-\mbox{Tr}_{\mbox{\tiny{A}}}(\rho_{\mbox{\tiny{A}}}\ln\rho_{\mbox{\tiny{A}}})$, where $\rho_{\mbox{\tiny{A}}}=\mbox{Tr}_{\tiny{\overline {\rm A}}}\rho$ is the reduced density matrix of the subsystem A.  This quantity has appeared in recent investigations in several domains including quantum field theory, condensed matter physics, quantum computing, and black hole physics. It is a measure of one's ignorance of the full system due to quantum entanglement between the degrees of freedom in the subsystem A and its complement  $\overline {\rm A}$.  
 
For  $d$+1-dimensional systems with local dynamics the entanglement entropy typically obeys an area law $S\sim A_{d-1}/\epsilon^{d-1}$ for $d\ge 2$,
where $A_{d-1}$ is the $d-1$-dimensional area of the boundary dividing the subsystem from its complement \cite{Srednicki:1993im}.
Various discussions of this area law have been made in the literature, including extensive study of bosonic systems in Ref.~\cite{Plenio} and
fermionic systems in Ref.~\cite{Gioev} (the latter can involve extra logarithmic factors).
%\footnote{The entropy for fermions often contains an extra logarithm $S\sim A_{d-1}/\epsilon^{d-1} \ln(A_{d-1}/\epsilon^{d-1})$.  See \cite{Swingle}.} 
%(reminiscent of the black hole thermal entropy law). 
Without further refinement the constant of proportionality is usually ill-defined, as it depends sensitively on an ultraviolet cutoff $\epsilon$. 
By contrast, the entanglement entropy of 1+1-dimensional systems is well-defined, since the $\epsilon$ dependence is only logarithmic.  For example, the entanglement entropy between a pair of half-spaces of a 1+1-dimensional conformal field theory at correlation length $\xi$ was shown to be $S=c/6\,\ln\,\xi/\epsilon$, where $c$ is the central charge of the conformal field theory \cite{Callan:1994py}; here rescaling of $\epsilon$ does not alter the coefficient of $\ln \xi$.   
This raises the challenge, to define and calculate unambiguous, cutoff-independent contributions to the entanglement entropy in $d\ge 2$ dimensions. 

In this Letter we do just that, albeit in the very special case of free field theory. 
We place the system in its ground state in order to isolate the entanglement entropy. Of course real systems are normally at finite temperature, which leads to a volume contribution to entropy, but this is not our focus. Instead our focus is toward gaining insight into novel quantum phenomena at zero temperature, such as quantum phase transitions.
For a free scalar field in $d$+1-dimensions at finite correlation length $\xi$ (i.e., mass $\mu=1/\xi$), we show that in addition to the divergent terms, such as $S\sim A_{d-1}/\epsilon^{d-1}$, there is also a finite area law contribution for general smooth geometries
% whose constant of proportionality depends on the correlation length:
\begin{eqnarray}
\Delta S \amp=\amp\gamma_d\,\frac{A_{d-1}}{\xi^{d-1}}\ln\frac{\xi}{\epsilon},\,\,\,\,
\mbox{for}\,\,\, d \,\,\, \mbox{odd},\nonumber\\
\Delta S \amp=\amp\gamma_d\,\frac{A_{d-1}}{\xi^{d-1}},\,\,\,\,\,\,\,\,\,\,\,\,\,\,\,
\mbox{for}\,\,\, d \,\,\, \mbox{even},
\label{AreaLaw}\end{eqnarray}
where $\gamma_d\equiv (-1)^{\frac{d-1}{2}}[6\,(4\pi)^{\frac{d-1}{2}}((d-1)/2)!]^{-1}$ for $d$ odd
and $\gamma_d\equiv(-1)^{d/2}[12\,(2\pi)^{(d-2)/2}(d-1)!!]^{-1}$ for $d$ even.  
For a waveguide geometry with specified boundary conditions (see Fig.~\ref{Waveguide}, left panel) we define and calculate additional power law corrections using heat kernel methods.   Those methods allow us to express the entropy as an expansion in terms of the geometric properties of the waveguide's cross-section.  We also consider a massless field ($\xi\to\infty$) and
%, using similar methods, 
define and calculate unambiguous finite terms for the interval in a waveguide (see Fig.~\ref{Waveguide}, right panel).

The area law for general smooth geometries in the massive case, as well as the waveguide expansion for both the massive and massless cases, extend the results of Ref.~\cite{Callan:1994py}.

%\bigskip 

{\em Heat Kernel Method}:  The replica trick is a powerful method for computing the entanglement entropy.   Since the logarithm that appears in the definition of that quantity is awkward to compute directly, one exploits the identity $S=-\mbox{Tr}_{\mbox{\tiny{A}}}(\rho_{\mbox{\tiny{A}}}\ln\rho_{\mbox{\tiny{A}}})
=\left(-\frac{d}{dn}+1\right)\ln\mbox{Tr}\rho_{\mbox{\tiny{A}}}^n |_{n=1}$ and the strategy to compute $\mbox{Tr}\rho_{\mbox{\tiny{A}}}^n$ for integer $n$ and use analytic continuation.   (A similar analytic continuation is known to fail for spin glasses \cite{parisi}, and we suspect that it can fail for the entanglement entropy in complicated quantum field theories, but it should be safe for the very simple field theories considered here.)     
Consider, for example, a field theory in 1 spatial dimension.  In that case the quantity 
$\mbox{Tr}\rho_{\mbox{\tiny{A}}}^n$ is a trace over an $n$-sheeted Riemann surface with cut along the subsystem of interest A.  If the subsystem A is a half-space, then as explained in Ref.~\cite{Callan:1994py} $\mbox{Tr}\rho_{\mbox{\tiny{A}}}^n$ can be identified with the partition function $Z_\delta$ on a cone of deficit angle $\delta=2\pi(1-n)$, and the entropy can be recast as
\begin{eqnarray}
S= \left(2\pi\frac{d}{d\delta}+1\right)\ln Z_\delta\Big{|}_{\delta=0}
\end{eqnarray}
which can be calculated analytically.

%{\em Finite Correlation Length.} ---
%Let's follow the heat kernel method developed in Ref.~\cite{Callan:1994py}.
Now consider the waveguide geometry shown in Fig.~\ref{Waveguide}, left panel.
The field lives in the bulk interior of the waveguide, satisfying boundary conditions on its surface.
For this geometry we formulate a Euclidean field theory on the space $C_\delta\times\mathcal{M}_{d-1}$,
where $C_\delta$ is a 2-dimensional cone of radius $R$ (infrared cutoff) and deficit angle $\delta$,
and $\mathcal{M}_{d-1}$ is the $(d-1)$-dimensional cross-section of the waveguide.
The cone's radius ($R\to\infty$) corresponds to the physical region in space we are tracing over
and the angular direction is associated with a geometric ``temperature'' (imaginary time) direction in the Euclidean path integral. Let $Z_\delta$ be the partition function for a field in its ground state defined on this space.
For a free scalar field of inverse mass $\xi$, the partition function is Gaussian 
\begin{eqnarray}
\ln Z_\delta = -\frac{1}{2}\ln\det\left(-\Delta+\xi^{-2}\right),
\end{eqnarray}
where $\Delta$ is the Laplacian satisfying the appropriate boundary conditions on $C_\delta\times\mathcal{M}_{d-1}$.

Now let us introduce the heat kernel for the Laplacian operator
%\begin{eqnarray} 
$\zeta(t)\equiv \mbox{tr}\left(e^{t\Delta}\right)$.
%\end{eqnarray}
The trace is defined by imposing Dirichlet or Neumann boundary conditions on the waveguide $\partial\mathcal{M}_{d-1}$ and Dirichlet boundary conditions on $\partial C_{\delta}$.
This allows us to rewrite the partition function $Z_\delta$ and hence the entropy $S$ in terms
of the heat kernel:
\begin{eqnarray}
S=\frac{1}{2}\int_0^\infty\frac{dt}{t}\left(2\pi\frac{d}{d\delta}+1\right)\zeta(t)e^{-t/\xi^2}\Big{|}_{\delta=0}.
\label{Sdelta}\end{eqnarray}

Since the manifold for the Euclidean field theory for the waveguide
is a direct product $C_\delta\times\mathcal{M}_{d-1}$, the heat kernel factorizes as
%\begin{eqnarray}
$\zeta(t)=\zeta_\delta(t)\,\zeta_{d-1}(t)$.
%\end{eqnarray}
Thus our problem simplifies into that of obtaining expansions for two separate heat kernels:
One for the 2-dimensional cone $\zeta_\delta(t)$; and the other for
the $(d-1)$-dimensional cross-section of the waveguide $\zeta_{d-1}(t)$.
The heat kernel for the cone has the form \cite{Alvarez:1982zi}
\begin{eqnarray}
\zeta_\delta(t) = \frac{1}{12}\left(\frac{2\pi}{2\pi-\delta}-\frac{2\pi-\delta}{2\pi}\right)+\ldots
\label{heatcone}\end{eqnarray}
where the dots represent terms that either are annihilated by the $2\pi\frac{d}{d\delta}+1|_{\delta=0}$ operator
or else vanish in the $R\to\infty$ limit, and therefore do not contribute to the entropy.  We thereby obtain 
\begin{eqnarray}
S=\frac{1}{12}\int_0^\infty\frac{dt}{t}\zeta_{d-1}(t)e^{-t/\xi^2}
\label{EntropyHeat}\end{eqnarray}
for the entanglement entropy for waveguide geometry in $d$ spatial dimensions traced over half-space.
Evidently the entropy is determined by the geometry of the waveguide cross-section, through its heat kernel $\zeta_{d-1}(t)$.

%\bigskip

{\em Waveguide Cross-Section}:
%Lets denote the eigenvalues of the Laplacian $\Delta$ by $-\mathcal{E}$.
%\zeta_{d-1}(t) = \int_{-\infty}^\infty d\mathcal{E}\rho_{d-1}(\mathcal{E})e^{-t\,\mathcal{E}}
%\end{eqnarray}
%\begin{eqnarray}
%\rho_0(\mathcal{E}) \amp = \amp \delta(\mathcal{E}),\nonumber\\
%\rho_1(\mathcal{E}) \amp = \amp \frac{a}{2\pi\sqrt{\mathcal{E}}}\Theta(\mathcal{E})+\frac{\eta}{2}\delta(\mathcal{E}),\nonumber\\
%\rho_2(\mathcal{E}) \amp = \amp \left(\frac{A}{2\pi}+\frac{\eta\,P}{8\pi\sqrt{\mathcal{E}}}\right)\Theta(\mathcal{E})+
%\chi\delta(\mathcal{E}).
%\label{dosexp}\end{eqnarray}
The heat kernel for a closed domain satisfying either Dirichlet ($\eta=-1$) or Neumann ($\eta=+1$) boundary conditions in dimensions 0, 1, and 2 has the following small $t$ expansion \cite{Baltes}:
\begin{eqnarray}
\zeta_0(t) \amp = \amp 1,\nonumber\\
\zeta_1(t) \amp = \amp \frac{a}{2\sqrt{\pi\,t}}+\frac{\eta}{2}+\ldots,\nonumber\\
\zeta_2(t) \amp = \amp \frac{A}{4\,\pi\,t}+\frac{\eta\,P}{8\sqrt{\pi\,t}}+\chi+\ldots.
\label{HeatExp}\end{eqnarray}
Here $a$ is the cross-sectional length of a waveguide in 2-dimensions, while
$A$, $P$, and $\chi$ are the cross-sectional area, perimeter length, and integrated curvature 
of a waveguide in 3-dimensions, respectively. (This expansion is also of use in computations of the Casimir effect between two partitions in a waveguide, see Ref.~\cite{Hertzberg:2005pr}.) 
The curvature term for an arbitrary piecewise smooth 2-dimensional cross-section is given by
\begin{eqnarray}
\chi = \sum_i\frac{1}{24}\left(\frac{\pi}{\alpha_i}-\frac{\alpha_i}{\pi}\right)
+\sum_j\frac{1}{12\pi}\int_{\gamma_j}\kappa(\gamma_j)d\gamma_j,
\end{eqnarray}
where $\alpha_i$ is the interior angle of any sharp corners and $\kappa(\gamma_j)$ is the curvature of any smooth pieces. For example, $\chi=1/6$ for any smooth shape (such as a circle)
and $\chi=(n-1)/(n-2)/6$ for any $n$-sided polygonal (so $\chi=1/4$ for a square). This result differs from Ref.~\cite{Fursaev:2006ng} where the curvature piece was argued to be proportional to the number of corners in an arbitrary shape. 
%The result of Ref.~\cite{Fursaev:2006ng} is only correct for a square cross-section.

%\bigskip

{\em Regularization - Finite Terms}: Direct insertion of the heat kernel expansion into eq.~(\ref{EntropyHeat}) for the entanglement entropy leads to divergences as $t\to0^+$.   These divergences are associated with the behavior of the theory at arbitrarily short distances.  As is known, these lead to infinities that cannot be renormalized away: 
%\cite{'tHooft:1984re}; 
logarithmic in 1 dimension,
linear in 2 dimensions, and quadratic in 3 dimensions \cite{Srednicki:1993im}.
There are various ways to regulate the divergences.  For instance, we could impose a hard cutoff on the $t$ integral and integrate from $t=t_c=\epsilon^2$ to $t=\infty$, and find terms that only diverge in the $\epsilon\to0$ limit. 
Another procedure is to use Pauli-Villars regularization by subtracting off terms with $\mu$ replaced by $\Lambda$ and taking $\Lambda$ large. This is perhaps more appealing as it respects the underlying geometry.
However, either approach gives results containing the cutoff parameters $\epsilon$ or $\Lambda$.

Fortunately, by returning to eq.~(\ref{EntropyHeat}) we can identify cutoff independent dependence of the entropy on the inverse correlation length $\mu=1/\xi$. In general, the leading order behavior of the heat kernel as $t\to0$ is
\begin{eqnarray}
\zeta_{d-1}(t)=\frac{\alpha}{t^{(d-1)/2}}+\ldots
\label{Heatdim}\end{eqnarray}
where $\alpha=A_{d-1}/(4\pi)^{(d-1)/2}$ is a constant. Inserting this into eq.~(\ref{EntropyHeat}) reveals that
the integrand has the leading order behavior $\sim 1/t^{(d+1)/2}$; giving a divergence of order $d-1$ as $t\to 0^+$ with respect to a cutoff, say $\epsilon$, defined through $t_c=\epsilon^2$.  (For $d-1=0$ there is a logarithmic divergence.)   This singularity can be regulated by taking some number of partial derivatives of the entropy with respect to the correlation length $\xi$, as that procedure pulls down factors of $t$ from the exponential $\exp(-t/\xi^2)$.
In particular by taking 
\begin{eqnarray}
k\equiv\mbox{Floor}\left[\frac{d+1}{2}\right]
\end{eqnarray}
derivatives of $S$ with respect to $\xi^{-2}$ gives a manifestly finite integral whose value is independent of any cutoff (note $k=1,1,2$ for $d=1,2,3$).
Hence we define a dimensionless, cutoff-independent quantity through
\begin{eqnarray}
S_\xi \equiv (-\xi^{-2})^k\frac{\partial^{k}\!S}{\partial(\xi^{-2})^k}.
\end{eqnarray}
Using eqs.~(\ref{EntropyHeat}) and (\ref{HeatExp}) and integrating $t$, 
%in the domain $t\in(0,\infty)$, 
we obtain
\begin{eqnarray}
S_\xi \amp = \amp \frac{1}{12},
\,\,\,\,\,\,\,\,\,\,\,\,\,\,\,\,\,\,\,\,\,\,\,\,\,\,\,\,\,\,\,\,\,\,\,\,\,\,\,\,\,\,\,\,\,\,\,\,\,\,\,\,\,\,\,\,\,\,\,\,\,\,\,\,\,\,\,\,\,\mbox{for}\,\,\,d=1,\nonumber\\
S_\xi \amp = \amp \frac{1}{24}\frac{a}{\xi}+\frac{\eta}{24}+\ldots,
\,\,\,\,\,\,\,\,\,\,\,\,\,\,\,\,\,\,\,\,\,\,\,\,\,\,\,\,\,\,\,\,\,\,\,\mbox{for}\,\,\,d=2,\nonumber\\
S_\xi \amp = \amp \frac{1}{48\pi}\frac{A}{\xi^2}+\frac{\eta}{192}\frac{P}{\xi}+\frac{\chi}{12}+\ldots,
\,\,\,\,\,\mbox{for}\,\,\,d=3.
\label{Suni}\end{eqnarray}
%which is our first primary result.
%Note that this is manifestly finite and the coefficients are universal.
For $d=1$ this result is exact \cite{Callan:1994py}.
For $d\geq 2$ this expansion is only valid for $a\gg\xi$, where $a$ is a typical cross-sectional length.  %(See below for exact results in $d=2$ and a square cross-section in $d=3$.) 
The third term is topological for $d=3$.
Note that by taking the appropriate number of anti-derivatives it is straightforward to isolate cutoff-independent contributions to the entropy itself.
%(We report additional exact results in the arXiv version of this Letter.)
We have also computed some exact results. In $d=2$ and for a square cross-section of width $a$ in $d=3$, $\zeta_{d-1}(t)$ is known exactly.  The result for $S_\xi$ is \begin{eqnarray} S_\xi \amp = \amp \frac{a \coth(a/\xi)}{24\,\xi}+\frac{\eta}{24}, \,\,\,\,\,\,\,\,\,\,\,\,\,\,\,\,\,\,\,\,\,\,\,\,\,\,\,\,\,\,\,\,\,\,\,\,\,\,\,\,\,\,\,\,\,\,\,\,\,\,\,\,\,\,\mbox{for}\,\,\,d=2\nonumber\\ S_\xi \amp = \amp \frac{a^2}{48\pi\xi^2} \Big{[}1+2\,\frac{a}{\xi}\sum_{n,m}\,\!\!'  f_{n,m} K_1\!\left(2f_{n,m}a/\xi\right) \Big{]}\nonumber\\ \amp+\amp\frac{\eta\,a}{48\,\xi}\left[\coth(a/\xi)+\frac{a}{\xi}\,\mbox{csch}(a/\xi)\right] +\frac{1}{48}, \,\,\,\,\mbox{for}\,\,\,d=3\nonumber \label{Sappendix}\end{eqnarray} where $f_{n,m}\equiv \sqrt{n^2+m^2}$, the primed summation means $\{n,m\}\in\mathbb{Z}^2/\{0,0\}$, and $K_1$ is the modified Bessel function of the second kind of order 1. For $a\gg\xi$ we recover eq.~(\ref{Suni}) plus exponentially small corrections. (Note that for the square $A=a^2$, $P=4\,a$, $\chi=1/4$.)

%\bigskip

{\em General Geometries}:
Although the subleading terms in eq.~(\ref{Suni}) are specific to a waveguide geometry, the leading terms have a meaning for arbitrary geometries. In particular, for any boundary in 1 dimension we pick up a contribution of 1/12 to $S_\xi$, as is known \cite{Calabrese:2004eu}. For closed geometries in 2 dimensions, the leading contribution is $S_\xi = P/(24\xi)$, where $P$ is the perimeter length.  For closed geometries in 3 dimensions, the leading contribution is
$S_\xi = A/(48\pi\xi^2)$. By integrating up these results, we recover the $d=1,2,3$ cases %that we reported 
in  eq.~(\ref{AreaLaw}). Furthermore, using the heat kernel in arbitrary dimensions (\ref{Heatdim}) we recover the general result for arbitrary dimensions.
%which is a universal area term in 3 dimensions for arbitrary geometries.

This general result differs from estimates made in Section 7 of Ref.~\cite{Ryu:2006ef}, where the corresponding
term in the entropy did not appear.
We have checked our result numerically for the cases of spheres and cylinders, finding excellent agreement.
In fact our numerics suggests that the area term is the only polynomial contribution to $S_\xi$ for large $A/\xi^2$. We can understand that heuristically, as follows: In the regime $\xi\ll a$, where $a$ is a typical length scale of curvature of the boundary, the correlations required to feel the curvature are exponentially suppressed.  On the other hand if the boundary contains sharp corners, we expect power law corrections to appear. (For related discussion, see \cite{Casini:2006hu}.) 
We have verified this numerically for squares.  
%For smooth geometries, though, it appears that subleading terms appear only through local boundary constraints, which modify the spectrum, as we found for the waveguide. 
%and for possible corrections from excited states see \cite{Das:2005ah,Das:2007pa}).

%\bigskip

{\em Massless Case - Finite Interval}:  The previous expansion requires the field theory to be massive. 
Let us turn now to the critical case ($\xi\to\infty$). To use our strategy to define finite entropy quantities we need a length scale, which will now come from considering a finite interval of length $L$, as in  Fig.~\ref{Waveguide}, right panel.  We can define the cutoff-independent quantity:
\begin{eqnarray}
S_L\equiv L\frac{dS}{dL}.
\end{eqnarray}

The small $t$ heat kernel expansion of eq.~(\ref{heatcone}) is insufficient here because we must know the form of $\zeta_\delta(t)$ not only for $t\lesssim L^2$, but also for $t\gtrsim L^2$ where $t$ is large.
In general the full form of $\zeta_\delta(t)$ is difficult to calculate. However, we do not need $\zeta_\delta(t)$ for arbitrary $\delta$, but only the specific limit indicated in eq.~(\ref{Sdelta}).  There are powerful tools available for this, as we now explain.  The derivative of the entanglement entropy can be written in terms of an object defined for 2d conformal field theories known as the $c$-function, denoted $C$.  It is related to the inverse Laplace transform of $\frac{1}{12}(2\pi\frac{d}{d\delta}+1)\zeta_\delta(t)|_{\delta=0}$. 
Convolving with the transverse density of states, we have
\begin{eqnarray}
S_L = \int_0^\infty d\mathcal{E}\,C\big{(}L\sqrt{\mathcal{E}}\big{)}\rho_{d-1}(\mathcal{E}).
\end{eqnarray}
The $c$-function $C$ has been studied intensely, see Ref.~\cite{Casini:2004bw}.
It is known that $C(0)=1/3$ and that $C$ is monotonically decreasing.
Using the heat kernel expansion (\ref{HeatExp}), we can inverse Laplace transform to obtain an expansion for the density of states  $\rho_{d-1}(\mathcal{E})$. The quantity $S_L$ can then be expressed in terms of a few integrals of $C$, which have been computed numerically. We find
\begin{eqnarray}
S_L \amp = \amp \frac{1}{3}
,\,\,\,\,\,\,\,\,\,\,\,\,\,\,\,\,\,\,\,\,\,\,\,\,\,\,\,\,\,\,\,\,\,\,\,\,\,\,\,\,\,\,\,\,\,\,\,\,\,\,\,\,\,\,\,\,\,\,\,\,\,\,\,\,\,\,\,\mbox{for}\,\,\,\,d=1,\nonumber\\
S_L \amp = \amp k_1\frac{a}{L}+\frac{\eta}{6}+\ldots
,\,\,\,\,\,\,\,\,\,\,\,\,\,\,\,\,\,\,\,\,\,\,\,\,\,\,\,\,\,\,\,\,\,\mbox{for}\,\,\,\,d=2,\nonumber\\
S_L \amp = \amp k_2\frac{A}{L^2}+\frac{\eta\, k_1}{4}\frac{P}{L}+\frac{\chi}{3}+\ldots,
\,\,\,\,\,\mbox{for}\,\,\,\,d=3.
\end{eqnarray}
%which is our second primary result.
Here $k_1\equiv\frac{1}{\pi}\int_0^\infty dx\,C(x)$ and $k_2\equiv\frac{1}{2\pi}\int_0^\infty dx\,x\,C(x)$.
The numerical values are: $k_1\approx 0.04$ %$0.039$
and $k_2\approx 0.01$ \cite{Ryu:2006ef}. %$0.0098$
For $d\ge 2$ this expansion is valid for $a\gg L$, analogous to the expansion in (\ref{Suni}) which was valid for $a\gg \xi$.

%\bigskip

{\em Discussion}: 
%\begin{enumerate}
%\item  
(i) We have shown that arbitrary shaped domains have an area term, given by eq.~(\ref{AreaLaw}), with a cutoff independent
piece emerging after taking $k=\mbox{Floor}[(d+1)/2]$ derivatives of $S$ with respect to $\xi^{-2}$. 
%This cutoff independent quantity can be defined for a single domain A.  
For a waveguide geometry we used our construction to obtain an asymptotic expansion of the entropy for small values of the correlation length to cross-section width ratio.  For arbitrary smooth manifolds the leading order area law should be applicable.   
In contrast to the 1-dimensional case, with $S_\xi=1/12$, these higher-dimensional entropies can be large numerically.   
It would be interesting to extend our results to fields involving alternative dispersion relations, fermions, and interacting field theories. 
%\vspace{1mm}
%\item 

(ii) Measurement of entanglement entropy of the kind discussed above, in the massive, or non-critical, case, requires changing the correlation length $\xi$ in such a way that the microphysics is only weakly affected, and measuring the corresponding change in entropy  $\Delta S$ (leaving the %incalculable 
cutoff-dependent pieces, such as $A_{d-1}/\epsilon^{d-1}$, unaffected). 
Though fluctuations of the vacuum state of a relativistic QFT may not be directly measurable \cite{Piazza:2009zh}, we can turn to condensed matter systems. 
Consider, for example, magnetic media.
In the absence of an external magnetic field, there is a massless mode; but for an externally applied $B$-field, $\phi$ acquires an adjustable effective mass $\mu=1/\xi$. In the regime: $\epsilon^2\ll \xi^2\ll A$, where $\epsilon$ is the inter-spin spacing, the area law should be an adequate description.  Another possibility would be to work near, but not too near, a quantum phase transition; then the correlation length could be varied  in a controlled way.  

(iii) A proposal for obtaining the entanglement entropy experimentally by measuring current fluctuations in 1-dimensional electron systems has been presented in Ref.~\cite{Klich:2008un}. Extending such proposals to 3-dimensions would be of great interest.
%\end{enumerate}

%\bigskip
\vspace*{1mm}

{\em Acknowledgments}:
We would like to thank Carlos Santana, Brian Swingle, and Erik Tonni for helpful discussions. 
%We would like to thank Mustafa Amin, Carlos Santana, Brian Swingle, and Max Tegmark for helpful discussions. 
We thank the Department of Energy (D.O.E.) for support under cooperative research agreement DE-FC02-94ER40818 and M.P.H thanks the Kavli Foundation.\newline
\indent$^*$Electronic address: mphertz@mit.edu

\end{document}